%% file: main.tex
\documentclass{article}
\usepackage{authblk}
\usepackage{orcidlink}
\usepackage{natbib}
\usepackage{amsmath, amssymb, amsthm}
\usepackage{cleveref}
\usepackage{graphicx} 

\title{The Fourth-Root Complexity of Data Movement}

\author{Chen Ding\orcidlink{0000-0003-4968-6659}}
\affil{University of Rochester, Rochester, New York, USA}

\begin{document}

\maketitle

\begin{abstract}
    Time complexity typically assumes $O(1)$ cost per data access.  This paper presents an analysis based on an abstract memory hierarchy.  For a common class of applications, it shows that the data-access cost scales with the fourth root of data size, that is, as data size $N$ increases, the cost of each access increases at the rate of $N^\frac{1}{4}$.

    While the analysis does not predict performance, it predicts scalability.  Specifically, the paper provides a precise analysis that shows the constant-factor difference between cases where the miss ratio follows a power law versus an exponential decay.
    
\end{abstract}

\section{Introduction}

Scaling analysis in computer science characterizes how performance or resource demands change with the size of the problem. Instead of empirically measuring every cache size and data size combination, the analysis identifies the general trend---such as $N$, $\log N$, or $N^{1/4}$---that governs the dominant system behavior at large inputs.

Time complexity analysis typically ignores data operations, implicitly assuming a constant cost at each data access.  
A recent cost model is called Data Movement Distance (DMD), which quantifies the cost of memory access by the distance of data movement in an abstract cache hierarchy~\citep{Snyder+:HIPS21,Smith+:ICS22}.

This paper analyzes the average DMD per data access and how it scales.  When provisioning the cache capacity, a well-known rule-of-thumb is \emph{$\sqrt{2}$ rule}.  As stated in \citet{Hartstein+:JILP08}: \emph{if you double the problem size, you need to multiply the cache size by $\sqrt{2}$ to maintain the same cache hit ratio}.  The rule had been empirically validated across workloads in database systems (scaling buffer pool sizes), file system caches, and Web caching. 

The analysis shows that DMD per access increases with data size $N$ as a biquadratic root function, in particular, the multiplicative factor of this scaling under two types of miss ratio function over cache size: the power-law decay and the exponential decay.

The change in the miss ratio depends on the parameters of the power law and exponential decay.  The analysis shows not only how the two decay functions affect DMD per access but also how the effect is determined by their parameters.  It shows mathematically how the decay functions and their parameters contribute to increasing or decreasing the average data access cost.

The scaling of the memory cost has been studied extensively in the past.  I/O complexity shows how the number of memory blocks transferred between slow and fast memory changes with the cache size and the problem size~\citep{HongK:STOC81,Frigo+:FOCS99}. However, it does not distinguish access costs beyond the binary hit-and-miss distinction. Uniform Memory Hierarchy (UMH) models the combined effect of memory layers~\citep{Alpern+:UMH94}. However, the model is general, so performance scaling depends on UMH parameters.  In comparison, this paper focuses on the memory cost based on a single machine model.

As a measure of the cost of data access, DMD is a theoretical model that captures locality in a simple yet general form. It abstracts away machine-specific details—such as cache levels, sizes, management policies, latency costs, and bandwidth limitations—leaving only the hierarchical structure of memory and the variation in data movement costs that depend on the locality of a computation. When presenting the PRAM (parallel random access machine) model for parallel computing, \citet{Vishkin:ICALP91} argues that without an idealized model that strips away the complexities of specific hardware designs, it would be difficult to develop general-purpose parallel algorithms. Following the same line of reasoning, the DMD complexity model offers a framework for general-purpose, memory-efficient computing.




\section{Complexity Analysis of Data Access Cost}

\subsection{Per Access Data Movement Distance (DMD)}
\label{sec:dmc}

Modern computers all use a memory hierarchy but have vastly different cache architectures, including the number of cache layers, the size of the caches, and their replacement algorithms.   
Data Movement Distance (DMD) is an abstract measure of memory cost~\citep{Smith+:ICS22,Snyder+:HIPS21}.  It assumes an abstract cache hierarchy, where (1) each layer adds one cache block to the previous layer, and (2) the access cost at each layer is the square-root of the cumulative cache size at this layer (assuming two-dimensional memory).  

In the abstract cache level $x$, DMD is the cost of moving the data when an access happens at this level.  In two-dimensional memory, DMD is defined as $\sqrt{x}$.  In physical space, the data size is proportional to the area of a memory cell, while DMD is proportional to its perimeter.  Therefore, the unit of DMD is the square-root of one unit of data.  

The \emph{data-access complexity} is the sum of DMD of all data access.  In this paper, we use the average DMD, which is the per-access DMD.  For brevity, DMD and per-access DMD are used interchangeably in this paper. 

\medskip

In continuous math, we assume that the miss ratio of a program is a continuous real function $f(x)$ for $x\ge 0$, where the domain of $x$ is a continuous approximation of the integral cache size.  DMD is an integration as follows:
\[
\mathcal{D}(N) = \int_0^\infty -f'(x,N) \sqrt{x} dx.
\]
where $f(x,N)$ is the miss ratio at cache level $x$ and data size $N$.  At each level $x$, the cost is the negative of the slope of the miss ratio function times the square root of $x$, which is the cumulative cache size at and above this level.  

The negative slope, $-f'(x,N)$, can be interpreted as the marginal benefit of adding cache at size $x$, or its ``benefit density.''  The weighted sum of marginal benefits from all caches is the average cost of access in the abstract cache hierarchy.

\subsection{Bounded Power Law}

Define the miss ratio function as a function of cache size \( x \) (without problem size \( N \) ) by
\[
f(x) = \frac{1}{(1 + a x)^\alpha}, \qquad a > 0,\ \alpha > 0.
\]
This function satisfies \( f(0) = 1 \) (zero cache means all misses) and decays as a power law for large \( x \) since \( f(x) \sim (a x)^{-\alpha} \).\footnote{A \emph{pure power law} would be:
\[
H_{\text{pure}}(z) = z^{-\alpha}, \qquad z > 0.
\]
The function \( H(z) = \dfrac{1}{(1+az)^\alpha} \) is a \emph{smooth, bounded power-law}.  For large \( z \):
\[
H(z) = \frac{1}{(1+az)^\alpha} \sim \frac{1}{az^\alpha} = (az)^{-\alpha}.
\]
Thus \( H(z) \) exhibits \emph{power-law decay} with exponent \( \alpha \).
For small \( z \):
\[
H(z) \to 1 \quad \text{as} \quad z \to 0^+.
\]
This is realistic for miss ratios (1 at zero cache).
}

Let \( g(x) \) be the marginal DMD cost at cache size $x$:
\[
g(x) = -f'(x) \sqrt{x}.
\]

Compute the derivative of \( f(x) \). Since \( f(x) = (1 + ax)^{-\alpha} \), the chain rule gives
\[
f'(x) = -\alpha a (1 + ax)^{-\alpha - 1}.
\]

Substitute \( f'(x) \) into the definition of \( g(x) \):
\[
g(x) = -[-\alpha a (1 + ax)^{-\alpha - 1}] \sqrt{x}
= \alpha a \sqrt{x} \, (1 + ax)^{-\alpha - 1}.
\]

Set up the integral of \( g(x) \) from \( 0 \) to \( \infty \):
\[
I = \int_0^\infty g(x) \, dx = \alpha a \int_0^\infty \sqrt{x} \, (1 + a x)^{-\alpha - 1} \, dx.
\]

Make the substitution \( u = a x \). Then \( x = u/a \), \( dx = du/a \), and \( \sqrt{x} = u^{1/2} / a^{1/2} \). Substituting these into the integral gives
\[
I = \alpha a \int_0^\infty \frac{u^{1/2}}{a^{1/2}} (1 + u)^{-\alpha - 1} \cdot \frac{du}{a}.
\]

Simplify the factors of \( a \). We have \( a \cdot a^{-1/2} \cdot a^{-1} = a^{1 - 1/2 - 1} = a^{-1/2} \). Thus the integral becomes
\[
I = \alpha a^{-1/2} \int_0^\infty u^{1/2} (1 + u)^{-\alpha - 1} \, du.
\]

Use the standard Beta function integral, which states that for \( p > 0 \) and \( q > 0 \),
\[
\int_0^\infty \frac{u^{p-1}}{(1+u)^{p+q}} \, du = B(p, q) = \frac{\Gamma(p) \Gamma(q)}{\Gamma(p+q)}.
\]
In our integral, \( u^{1/2} = u^{p-1} \) so \( p - 1 = 1/2 \), hence \( p = 3/2 \). The denominator exponent is \( \alpha + 1 = p + q \), so \( q = \alpha + 1 - 3/2 = \alpha - 1/2 \). Therefore,
\[
\int_0^\infty u^{1/2} (1+u)^{-\alpha - 1} \, du
= B\!\left( \frac32,\ \alpha - \frac12 \right).
\]
Convergence requires \( q > 0 \), i.e., \( \alpha > 1/2 \).

Express the Beta function in terms of Gamma functions:
\[
B\!\left( \frac32,\ \alpha - \frac12 \right)
= \frac{ \Gamma(3/2) \, \Gamma(\alpha - 1/2) }{ \Gamma(\alpha + 1) }.
\]
Using \( \Gamma(3/2) = \frac12 \Gamma(1/2) = \sqrt{\pi}/2 \) and \( \Gamma(\alpha + 1) = \alpha \Gamma(\alpha) \), we get
\[
\int_0^\infty u^{1/2} (1+u)^{-\alpha - 1} \, du
= \frac{ \frac{\sqrt{\pi}}{2} \, \Gamma(\alpha - 1/2) }{ \alpha \, \Gamma(\alpha) }.
\]

Substitute this result back into the expression for \( I \):
\[
I = \alpha a^{-1/2} \cdot \frac{ \frac{\sqrt{\pi}}{2} \, \Gamma(\alpha - 1/2) }{ \alpha \, \Gamma(\alpha) }
= \frac{\sqrt{\pi}}{2\sqrt{a}} \cdot \frac{ \Gamma(\alpha - 1/2) }{ \Gamma(\alpha) }.
\]

Thus the final closed-form result is
\[
\boxed{ \int_0^\infty g(x) \, dx = \frac{\sqrt{\pi}}{2\sqrt{a}} \cdot \frac{ \Gamma(\alpha - 1/2) }{ \Gamma(\alpha) } }, \qquad \alpha > \frac12.
\]

Special cases include:
\[
\alpha = 1: \quad I = \frac{\pi}{2\sqrt{a}},
\]
\[
\alpha = 3/2: \quad I = \frac{1}{\sqrt{a}},
\]
\[
\alpha = 2: \quad I = \frac{\pi}{4\sqrt{a}}.
\]
For \( \alpha = 3/4 \) (which is less than 1), we get
\[
I = \frac{\sqrt{\pi}}{2\sqrt{a}} \cdot \frac{ \Gamma(1/4) }{ \Gamma(3/4) },
\]
which is a closed form in terms of Gamma functions but not elementary.

As \( \alpha \to \infty \), using Stirling's approximation \( \Gamma(\alpha - 1/2)/\Gamma(\alpha) \sim \alpha^{-1/2} \), we get \( I \to 0 \). As \( \alpha \to (1/2)^+ \), \( \Gamma(\alpha - 1/2) \to \infty \), so \( I \to \infty \), corresponding to the logarithmic divergence at the critical exponent.


\subsection{Streched Exponential Decay}

Define the miss ratio function as a function of cache size \( x \) for fixed problem size \( N \) by
\[
f(x) = e^{-(a x)^\beta}, \qquad a > 0,\ \beta > 0.
\]
This function satisfies \( f(0) = 1 \) (zero cache means all misses) and decays exponentially for large \( x \) since \( f(x) \sim e^{-(a x)^\beta} \). For \( \beta = 1 \), this reduces to the simple exponential \( f(x) = e^{-a x} \).

Let \( g(x) \) be the marginal DMD cost at cache size $x$:
\[
g(x) = -f'(x) \sqrt{x}.
\]

Compute the derivative of \( f(x) \). Since \( f(x) = e^{-(a x)^\beta} \), the chain rule gives
\[
f'(x) = -\beta a^\beta x^{\beta-1} e^{-(a x)^\beta}.
\]

Substitute \( f'(x) \) into the definition of \( g(x) \):
\[
g(x) = -[-\beta a^\beta x^{\beta-1} e^{-(a x)^\beta}] \sqrt{x}
= \beta a^\beta x^{\beta - 1/2} e^{-(a x)^\beta}.
\]

Set up the integral of \( g(x) \) from \( 0 \) to \( \infty \):
\[
I = \int_0^\infty g(x) \, dx = \beta a^\beta \int_0^\infty x^{\beta - 1/2} e^{-(a x)^\beta} \, dx.
\]

Make the substitution \( u = (a x)^\beta \). Then:
\[
x = \frac{u^{1/\beta}}{a}, \qquad dx = \frac{1}{\beta a} u^{1/\beta - 1} du, \qquad x^{\beta - 1/2} = \frac{u^{1 - 1/(2\beta)}}{a^{\beta - 1/2}}.
\]

Substitute these into the integral:
\[
I = \beta a^\beta \int_0^\infty \frac{u^{1 - 1/(2\beta)}}{a^{\beta - 1/2}} e^{-u} \cdot \frac{1}{\beta a} u^{1/\beta - 1} du.
\]

Simplify the factors of \( a \). We have:
\[
a^\beta \cdot a^{-(\beta - 1/2)} \cdot a^{-1} = a^{\beta - \beta + 1/2 - 1} = a^{-1/2}.
\]

Simplify the powers of \( u \):
\[
u^{1 - 1/(2\beta)} \cdot u^{1/\beta - 1} = u^{1/\beta - 1/(2\beta)} = u^{1/(2\beta)}.
\]

Thus the integral becomes:
\[
I = a^{-1/2} \int_0^\infty u^{1/(2\beta)} e^{-u} \, du.
\]

The integral is the Gamma function:
\[
\int_0^\infty u^{1/(2\beta)} e^{-u} \, du = \Gamma\!\left( 1 + \frac{1}{2\beta} \right).
\]

Using the recurrence relation \( \Gamma(1 + z) = z \Gamma(z) \) with \( z = 1/(2\beta) \):
\[
\Gamma\!\left( 1 + \frac{1}{2\beta} \right) = \frac{1}{2\beta} \Gamma\!\left( \frac{1}{2\beta} \right).
\]

Substitute this back into the expression for \( I \):
\[
I = a^{-1/2} \cdot \frac{1}{2\beta} \Gamma\!\left( \frac{1}{2\beta} \right).
\]

Thus the final closed-form result is
\[
\boxed{ \int_0^\infty g(x) \, dx = \frac{1}{2\beta\sqrt{a}} \, \Gamma\!\left( \frac{1}{2\beta} \right) }, \qquad \beta > 0.
\]

Special cases include:
\[
\beta = 1: \quad I = \frac{1}{2\sqrt{a}} \Gamma(1/2) = \frac{\sqrt{\pi}}{2\sqrt{a}},
\]
which is the simple exponential result.
\[
\beta = 2: \quad I = \frac{1}{4\sqrt{a}} \Gamma(1/4).
\]
\[
\beta = 1/2: \quad I = \frac{1}{\sqrt{a}} \Gamma(1) = \frac{1}{\sqrt{a}}.
\]

As \( \beta \to \infty \), using the asymptotic \( \Gamma(1/(2\beta)) \sim 2\beta \) for large \( \beta \), we get:
\[
I \sim \frac{1}{2\beta\sqrt{a}} \cdot 2\beta = \frac{1}{\sqrt{a}}.
\]
This corresponds to the step function limit where \( f(x) \) approaches 1 for \( x < 1/a \) and 0 for \( x > 1/a \).

As \( \beta \to 0^+ \), \( \Gamma(1/(2\beta)) \to \infty \), so \( I \to \infty \), corresponding to extremely slow decay.

\subsection{The $\sqrt{2}$ Rule of Cache-Size Scaling}
\label{sec:sqrt2}

When provisioning the cache capacity, a well-known rule-of-thumb is \emph{$\sqrt{2}$ rule}.  As stated in \citet{Hartstein+:JILP08}: \emph{if you double the problem size, you need to multiply the cache size by $\sqrt{2}$ to maintain the same cache hit ratio}.  The authors mentioned that this rule had been empirically validated across workloads in database systems (scaling buffer pool sizes), file system caches, and Web caching. 

\paragraph{A functional form of the \(\sqrt{2}\) rule}

The \(\sqrt{2}\) rule states: doubling the problem size \(N\) requires multiplying the cache size \(C\) by \(\sqrt{2}\) to maintain the same miss ratio:
\[
m(C, N) = m(\sqrt{2}\, C,\; 2N).
\]

Assume the miss ratio depends only on the ratio \(C / N^\beta\) for some \(\beta\). Let
\[
m(C, N) = H\!\left( \frac{C}{N^\beta} \right).
\]

Applying the rule:
\[
m(\sqrt{2}C, 2N) = H\!\left( \frac{\sqrt{2}C}{(2N)^\beta} \right)
= H\!\left( \frac{\sqrt{2}}{2^\beta} \cdot \frac{C}{N^\beta} \right).
\]

Equality with \(H(C/N^\beta)\) for all \(C,N\) implies:
\[
\frac{\sqrt{2}}{2^\beta} = 1 \quad \Rightarrow \quad 2^{1/2} = 2^\beta \quad \Rightarrow \quad \beta = \frac12.
\]

Thus:
\[
\boxed{m(C, N) = H\!\left( \frac{C}{\sqrt{N}} \right)}.
\]

\paragraph{Define \(f(x)\) for the $\sqrt{2}$ rule}

Fix \(N\) and let \(x = C\) (cache size). Define:
\[
f(x) = m(x, N) = H\!\left( \frac{x}{\sqrt{N}} \right).
\]
Having a single variable input $z=\frac{x}{\sqrt{N}}$, $H$ captures the cache-size scaling.  Any miss-ratio function can be $H$ as long as $H(0)=1$ and $H(\infty)=0$, except that the parameter $z$ is not the actual cache size, the cache size is $x = \sqrt{N} z$.  The miss-ratio function is \[
f(x) = H(\frac{x}{\sqrt{N}}).
\]

\paragraph{Basic DMD Scaling}

The $\sqrt{2}$ scaling rule is sufficient to derive the basic scaling of DMD. If a given access pattern has a DMD of $d$, then it is a cache hit for sizes $C \ge d^2$ and a miss for $C < d^2$. Under the cache-size scaling rule, doubling the data size shifts the hit threshold to $\sqrt{2}C$: the pattern becomes a hit for $C \ge \sqrt{2}\,d^2$ and a miss below that. Consequently, the new DMD becomes $\sqrt[4]{2}\,d$, that is, the fourth root.

More generally, to preserve the same miss ratio under the cache-size scaling rule, the cache size must scale proportionally to $\sqrt{N}$, where $N$ is the data size. This implies that the DMD scales as $N^{1/4}$. We refer to this as the \emph{basic DMD scaling}, as it determines the exponent but leaves the multiplicative constant unspecified. In what follows, we provide a precise analysis of the coefficient of $N^{1/4}$ in this scaling relation.




\paragraph{Precise DMD Scaling}

Recall the previous results:
\[
\text{Bounded power law: } \int_0^\infty g(x) \, dx = \frac{\sqrt{\pi}}{2\sqrt{a}} \cdot \frac{ \Gamma(\alpha - 1/2) }{ \Gamma(\alpha) }, \qquad \alpha > 1/2.
\]
\[
\text{Stretched exponential: } \int_0^\infty g(x) \, dx = \frac{1}{2\beta\sqrt{a}} \, \Gamma\!\left( \frac{1}{2\beta} \right), \qquad \beta > 0.
\]
The \( \sqrt{2} \) rule required \( f(x) = H(x/\sqrt{N}) \).  Substitute \( a = 1/\sqrt{N} \). Both scale as \( 1/\sqrt{a} \), which becomes \( N^{1/4} \).

\[
\text{Bounded power law: } \int_0^\infty g(x) \, dx = \frac{\sqrt{\pi}}{2} \cdot N^{1/4} \cdot \frac{ \Gamma(\alpha - 1/2) }{ \Gamma(\alpha) } , \qquad \alpha > 1/2.
\]
\[
\text{Stretched exponential: } \int_0^\infty g(x) \, dx = \frac{N^{1/4}}{2\beta} \, \Gamma\!\left( \frac{1}{2\beta} \right), \qquad \beta > 0.
\]

Both forms yield the same \( N^{1/4} \) scaling, but with different constants determined by the shape parameters \( \alpha \) and \( \beta \).
In Bounded Power Law, as $\alpha$ increases, the locality improves due to greater degrees of data reuse.  The per-access DMD quantifies this change of locality.  As $\alpha$ increases from $\frac{3}{4}$ to 2, the coefficient of DMD decreases from about 2.6 to $\frac{\pi}{4}\approx 0.79$
\begin{align*}
\alpha &= 3/4: & \frac{\sqrt{\pi}}{2} N^{1/4} \cdot \frac{\Gamma(1/4)}{\Gamma(3/4)} \approx 2.622 \, N^{1/4}.\\
\alpha &= 1: & \frac{\pi}{2} N^{1/4}. \\
\alpha &= 3/2: & N^{1/4}. \\
\alpha &= 2: & \frac{\pi}{4} N^{1/4}. \\
\end{align*}
In Stretched Exponential:
\begin{align*}
\beta &= 1/2: & \frac{N^{1/4}}{1} \Gamma(1) = N^{1/4}.\\
\beta &= 1: & \frac{\sqrt{\pi}}{2} N^{1/4}. \\
\beta &= 2: & \frac{N^{1/4}}{4} \Gamma(1/4). \\
\end{align*}

For the bounded power law with \( \alpha = 3/2 \), and the stretched exponential with \( \beta = 1/2 \), we get \( \int g = N^{1/4} \).

\subsection{A More General Case}

Instead of $\sqrt{2}$ scaling which means $f(x)=H(ax)$ for a specific $a$, we define 
\[
f(x) = H(ax)
\]
Compute:
\[
f'(x) = a H'(a x).
\]
Therefore:
\[
g(x) = -a H'(a x) \sqrt{x}.
\]
Integrate \(g(x)\) from \(0\) to \(\infty\).  
We want:
\[
I = \int_0^\infty g(x) \, dx = \int_0^\infty \left[ -a \sqrt{x} \, H'(a x) \right] dx.
\]
Substitute \(u = a x\), so \(x = u/a\), \(dx = du/a\), and \(\sqrt{x} = \sqrt{u/a} = u^{1/2} a^{-1/2}\).
\[
g(x) dx = \left[ -a \cdot u^{1/2} a^{-1/2} \cdot H'(u) \right] \cdot \frac{du}{a} = -a^{-1/2} \, u^{1/2} H'(u) \, du.
\]
Thus:
\[
I = -a^{-1/2} \int_0^\infty u^{1/2} H'(u) \, du.
\]
Integration by parts.  
Let \(p = u^{1/2}\), \(dq = H'(u) du\). Then:
\[
dp = \frac12 u^{-1/2} du, \quad q = H(u).
\]
\[
\int_0^\infty u^{1/2} H'(u) du = \left[ u^{1/2} H(u) \right]_0^\infty - \int_0^\infty H(u) \cdot \frac12 u^{-1/2} du.
\]
Boundary terms:
\begin{itemize}
  \item As \(u \to \infty\): \(H(u) \to 0\) (miss ratio vanishes for large cache), and if \(H(u)\) decays faster than \(u^{-1/2}\), then \(u^{1/2} H(u) \to 0\).
  \item As \(u \to 0^+\): \(H(u) \to 1\) (miss ratio \(=1\) when cache size \(=0\)), so \(u^{1/2} H(u) \to 0\).
\end{itemize}
Thus the boundary term vanishes.  Therefore:
\[
\int_0^\infty u^{1/2} H'(u) du = -\frac12 \int_0^\infty H(u) u^{-1/2} du.
\]
Substitute back
\[
I = -a^{-1/2} \left( -\frac12 \int_0^\infty H(u) u^{-1/2} du \right)
= \boxed{ \frac12 a^{-1/2} \int_0^\infty H(u) u^{-1/2} du }.
\]

In the case of $\sqrt{2}$ scaling, \(a = 1/\sqrt{N}\), so \(a^{-1/2} = (1/\sqrt{N})^{-1/2} = N^{1/4}\).
Thus:
\[
\int_0^\infty g(x) \, dx = \frac12 N^{1/4} \int_0^\infty H(u) \, u^{-1/2} \, du.
\]

\subsection{N-Invariant DMD}

Per access DMD is a constant value when the miss ratio does not change with the problem size $N$.  Recall the previous results:
\[
\text{Bounded power law: } \int_0^\infty g(x) \, dx = \frac{\sqrt{\pi}}{2\sqrt{a}} \cdot \frac{ \Gamma(\alpha - 1/2) }{ \Gamma(\alpha) }, \qquad \alpha > 1/2.
\]
\[
\text{Stretched exponential: } \int_0^\infty g(x) \, dx = \frac{1}{2\beta\sqrt{a}} \, \Gamma\!\left( \frac{1}{2\beta} \right), \qquad \beta > 0.
\]

Instead of $a$ as a function of $N$, N-invariant DMD depends on constant values of $a$.  If we set $a=1$, the result DMD is the same as that of $a=\frac{1}{\sqrt{N}}$ in \Cref{sec:sqrt2} after removing the power term $N^\frac{1}{4}$.  In Bounded Power Law, as $\alpha$ increases, locality improves from greater degrees of data reuse.  The per-access DMD decreases from about 2.6 to $\frac{\pi}{4}\approx 0.79$.

\input{rel.tex}

\section{Summary}

This paper has mathematically derived the DMD per access for two classes of miss ratio decay functions, power law and exponential.  When the cache scaling follows the $\sqrt{2}$ rule, it shows that the DMD cost scales as a biquadratic-root function of $N$.  The derivation shows DMD result for all $\alpha > \frac{1}{2}$ in power-law decay and all $\beta$ in stretched exponential decay.

DMD complexity isolates the intrinsic cost of data access as a function of pure locality. Although it does not predict performance on any specific machine, it pinpoints which locality characteristics will scale gracefully in memory cost as problem sizes grow—independent of future hardware memory systems. Despite its idealized nature, it provides a rigorous foundation for general-purpose, memory-efficient computing.

\subsection*{Acknowledgements}

The author wishes to thank Yifan Zhu, Yanhui (Woody) Wu, and Skylar \linebreak Abruzese for their help with the presentation of the paper.
AI tools, primarily DeepSeek-V3, were used to aid in mathematical derivations and to suggest improvements in sentence- and paragraph-level editing and organization, with all outputs reviewed and validated by the author.  

\bibliographystyle{ACM-Reference-Format}
\bibliography{all}

\end{document}

%% file: rel.tex
\section{Related Work}
\label{sec:rel}

Time and space complexity are fundamental concepts in computing. Like them, DMD analysis is asymptotic: in a polynomial, the highest-power term dominates the rest. However, time complexity analysis employs Big-O notation, which discards multiplicative constants--DMD analysis does not. In Big-O, unspecified constants are treated as arbitrary and equivalent, whereas in this work, the constant factor is precisely derived by DMD analysis.


The classic model called Random Access Machine (RAM) assumes uniform memory: a program can read from and write to anywhere in memory in a single unit of time. The 
I/O complexity (or the external memory model) measures the number of memory blocks transferred between slow and fast memory~\citep{HongK:STOC81,Frigo+:FOCS99}, with many follow-up results of a collection of algorithms, including recently on dynamic programming~\citep{DeStefaniG:WADS2025}. The access cost is given by a big-O function $O(f(C,N))$, indicating the lower-bound number of data transfers. The I/O complexity does not distinguish access costs beyond the binary hit-and-miss distinction. However, since it shows the cache size scaling with the problem size, it can be used to derive the DMD result using similar derivations in this paper.

For regular loop nests, dependence analysis identifies data reuses and the distance measured in the number of loop iterations~\citep{AllenK:Book01,CarrK:TOPLAS94,Aho+:Book06}. 
Reuse distance equations were first to derive the reuse distance symbolically using Ehrhart Polynomials~\citep{Verdoolaege+:CASES04,BeylsD:JSA05}.  Polyhedral compiler analysis counts cache misses symbolically when given a cache size $C$~\citep{Bao+:POPL18,Pitchanathan+:PLDI24}.  Our work recently derives a close-form, precise miss count as a function of both the cache size $C$ and problem size $N$~\citep{Zhu+:arXiv26-2}, building on a new concept called imaginary reuses and using the derivation based on Denning recursion~\citep{DenningS:CACM72,Yuan+:TACO19}.  Unlike in I/O compilexity, the symbolic count from program analysis is precise rather than in big-O form.  

Stochastic workloads have long been studied for their locality characteristics. Our recent continuous-time model~\citep{Wang+:TOMPECS25} generalizes earlier Denning recursion techniques~\citep{DenningS:CACM72} and the work of~\citet{DanT:SIGMETRICS90}. Its cell model produces symbolic cache size and miss ratio expressions in the form of natural exponential functions. In a related study,~\citet{Smith+:MEMSYS21} used integration to approximate the cost of cache writebacks under Zipfian workloads. 

When a fully symbolic form is available—whether derived from program analysis or workload modeling—it can be directly applied, as demonstrated in this paper, to derive DMD results.  In addition, the symbolic miss ratio can be used to derive other measures of cache behavior in particular the average eviction time~\citep{Hu+:TOS18,Yuan+:TACO19} using a new Cache Identity Equation~\citep{Liu+:HIPS26}.

An early model for scaling analysis is the Uniform Memory Hierarchy (UMH) model~\citep{Alpern+:UMH94}.  In UMH, layered memory is defined by two parameters, the number of blocks at each layer and the increase in block size.  An algorithm is evaluated by the ratio between its complexity under uniform memory cost and the complexity under UMH.  An algorithm is \emph{communication bound} if the ratio is arbitrarily close to zero; otherwise, it is communication efficient.  The abstract cache hierarchy used in our analysis has a single block size.  The cost is defined by DMD, while the cost in UMH is defined by a transfer cost function, valued in the number of cycles that may be constant or proportional to the size of each data block~\citep{Alpern+:UMH94}. In UMH, the capacity of the lowest layer dominates the total memory.  The transfer cost function of a UMH layer similar to DMD would be the square root of the memory size of the layer. 


\citet{Snyder+:HIPS21} analyzed the DMC of repeated traversals on $m$ data items and showed that the DMD per-access is $\sqrt{m}$ for the cyclic traversal and $\frac{2}{3}\sqrt{m}$ for the sawtooth traversal. In both cases, the cache size scales linearly with $m$. Their difference  is the miss ratio, not the cache-size scaling exponent or coefficient. One might naturally expect that a higher miss ratio could be compensated for by increasing the cache size, effectively absorbing the difference into either a larger coefficient or exponent in the cache size scaling relation. However, the cyclic and sawtooth cases demonstrate that such a conversion is not always possible: their miss ratios cannot be made equal by adjusting the cache capacity.  Separate from DMD analysis, we have formulated monotonicity and proportionality of data access complexity~\citep{DingZ:MEMSYS25}.